\documentclass[mathleft]{an}
\usepackage{graphicx}
\usepackage{times}
\usepackage{xspace}
\usepackage{natbib}
\bibpunct{(}{)}{;}{a}{}{,}
\newcommand{\cloudy}{\texttt{CLOUDY}\xspace}
\newcommand{\isis}{\texttt{ISIS}\xspace}
\newcommand{\spex}{\texttt{SPEX}\xspace}
\newcommand{\xstar}{\texttt{XSTAR}\xspace}
\newcommand{\xstardb}{\texttt{XSTARDB}\xspace}

\newcommand{\cyg}{{Cyg~X-1}\xspace}
\newcommand{\fu}{{4U~1957$+$11}\xspace}
\newcommand{\pg}{{PG1211$+$143}\xspace}

\newcommand{\athena}{\textsl{Athena}\xspace}
\newcommand{\chandra}{\textsl{Chandra}\xspace}
\newcommand{\epic}{\textsl{EPIC-PN}\xspace}
\newcommand{\heg}{\textsl{HEG}\xspace}
\newcommand{\hetg}{\textsl{HETG}\xspace}
\newcommand{\hitomi}{\textsl{Hitomi}\xspace}
\newcommand{\integral}{\textsl{INTEGRAL}\xspace}
\newcommand{\letg}{\textsl{LETG}\xspace}
\newcommand{\meg}{\textsl{MEG}\xspace}
\newcommand{\nustar}{\textsl{NuSTAR}\xspace}
\newcommand{\rgs}{\textsl{RGS}\xspace}
\newcommand{\swift}{\textsl{Swift}\xspace}
\newcommand{\suz}{\textsl{Suzaku}\xspace}
\newcommand{\xmm}{\textsl{XMM-Newton}\xspace}
\newcommand{\xrs}{\textsl{X-ray Surveyor}\xspace}

\begin{document}

\Pagespan{789}{}
\Yearpublication{2017}%
\Yearsubmission{2016}%
\Month{11}%
\Volume{999}%
\Issue{88}%
\DOI{This.is/not.aDOI}%

\title{Leveraging High Resolution Spectra to Understand Black Hole Spectra}

\author{Michael A. Nowak\inst{1}\fnmsep\thanks{Corresponding author:
  \email{mnowak@space.mit.edu}\newline}
}
\titlerunning{Leveraging High Resolution Spectra}
\authorrunning{Michael A. Nowak}
\institute{Massachusetts Institute of Technology, Kavli Institute for Astrophysics, 
  Cambridge, MA 02139, USA}

\received{9 October 2016}
\accepted{1 December 2016}
\publonline{later}


\abstract{For the past 17 years, both \xmm and \chandra have brought
  the powerful combination of high spatial and spectral resolution to
  the study of black hole systems. Each of these attributes requires
  special consideration--- in comparison to lower spatial resolution,
  CCD quality spectra--- when modeling observations obtained by these
  spacecraft.  A good understanding of the high resolution spectra is
  in fact required to model properly lower resolution CCD spectra,
  with the Reflection Grating Spectrometer (\rgs) instrument on \xmm
  maintaining the highest ``figure of merit'' at soft X-ray energies
  for all missions flying or currently planned for the next
  decade. Thanks to its even higher spectral resolution, the use of
  \chandra-High Energy Transmission Gratings (\hetg), albeit with
  longer integration times, allows for one to bring further clarity to
  \rgs studies.  A further promising route for continued studies is
  the combination of high spectral resolution at soft X-rays, via \rgs
  and/or \hetg, with contemporaneous broadband coverage extending to
  hard X-rays (e.g., \nustar or \integral spectra).  Such studies
  offer special promise for answering fundamental questions about
  accretion in black hole systems; however, they have received only
  moderate consideration to date.  This may be due in part to the
  difficulty of analyzing high resolution spectra. In response, we
  must continue to develop software tools that make the analysis of
  high resolution X-ray spectra more accessible to the wider
  astrophysics community.}

\maketitle

\section{Introduction}\label{sec:intro}

Fundamental questions remain in the study of astrophysical black
holes.  What are the most robust techniques for measuring black hole
spin? (See the reviews of \citealt{middleton:16a,reynolds:03a}.)  How
does this spin relate to the formation of jets in these systems?  (See
\citealt{fender:16a}.)  What are the relationships between jets and
outflowing winds, and how much mass and accretion energy is
transported by the latter?  (See \citealt{ponti:12a}.)  What fraction
of the hard X-rays can be explained by jet emission, and what fraction
can be attributed to a corona?  (See \citealt{markoff:15a}.)  Does the
disk recede as black holes transit from high/soft states to low/hard
states?  (See the review by \citealt{done:07a}.) Answers to these
questions have been sought via multi-wavelength observations ranging
from radio wavelengths through hard X-ray energies, as many of the
above cited phenomena manifest themselves over broad energy bands.

The soft X-ray, however, holds special importance for a number of
components.  It is where the disk spectrum peaks.  It is crucial for
studies of wind properties.  It dominates the bolometric luminosity in
Galactic black hole ``high states''.  Arguably, the two most important
instruments for study of the soft X-ray spectra of black hole systems
have been the \xmm and \chandra satellites.  Both of these instruments
share the properties of having unparalleled spatial and spectral
resolution.  The latter is achieved via the Reflection Gratings
Spectrometer (\rgs; \citealt{herder:01a}) of \xmm, and by the Low
Energy Transmission Gratings (\letg; \citealt{brinkman:00a}) and High
Energy Transmission Gratings (\hetg; \citealt{canizares:05a}) of
\chandra.  The \hetg is comprised of the High Energy Gratings (\heg)
and Medium Energy Gratings (\meg).  Both the high spatial and spectral
resolution properties of these instruments pose unique challenges that
are often overlooked in analyses of black hole systems.  The high
spectral resolution properties especially have not been utilized as
often as they should or could be in multi-wavelength campaigns.

We present examples below from both stellar mass and supermassive
black holes.  We briefly discuss the synergy between \rgs and \hetg
high spectroscopic resolution studies.  We then discuss the state
of software for high resolution spectroscopic analysis. We end with
considerations of the current use of high resolution spectra in
multi-satellite, broadband X-ray campaigns.

\section{Low/High Resolution Comparisons}\label{sec:lowhi}

\begin{figure*}
\begin{center}
\includegraphics[width=0.45\textwidth,viewport=30 385 575 790]{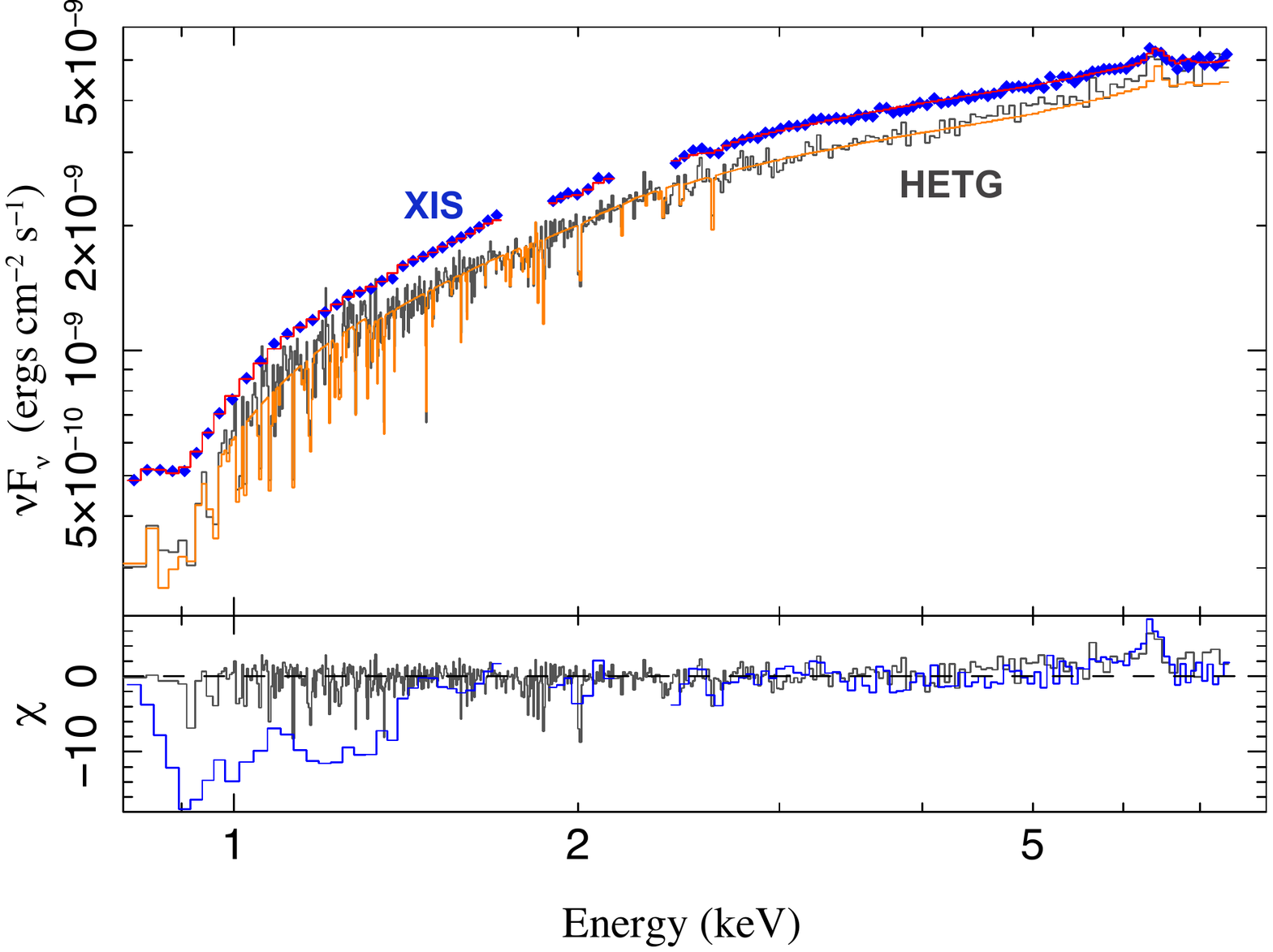} ~~~~~
\includegraphics[width=0.46\textwidth,viewport=30 385 578 790]{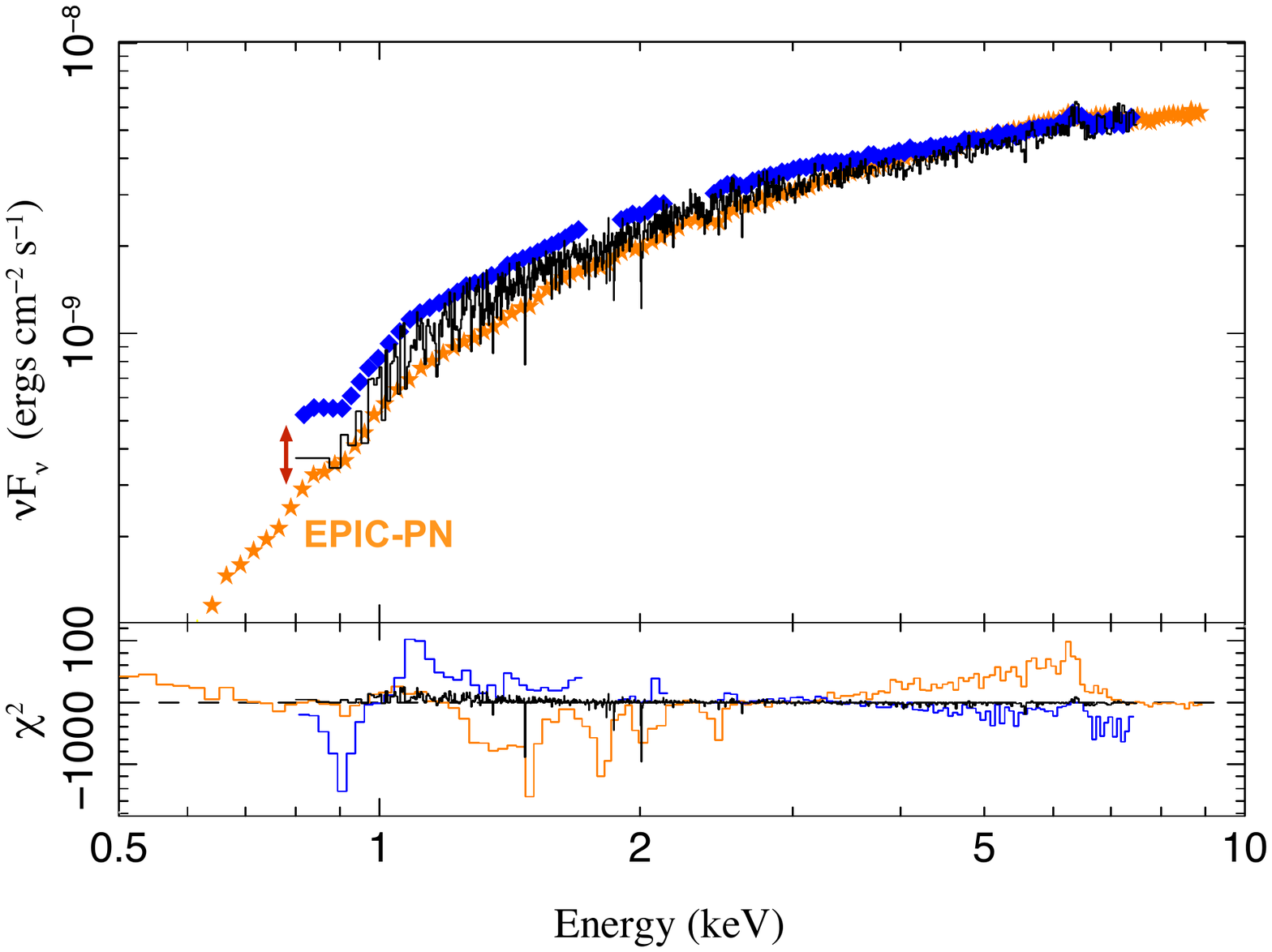}
\end{center}
\vspace{-0.1 in}
\caption{Left: Comparison of quasi-simultaneous flux-corrected
  \suz-XIS (blue diamonds) and \chandra-\hetg (grey histogram) spectra
  of a hard state of \cyg.  (Times of dips, see below, have been
  excluded.) The model (red and orange histograms, respectively)
  consists of a disk, powerlaw, and Fe\,K$\alpha$ emission line
  absorbed by both neutral and ionized emission.  Scattering of the
  soft X-ray photons out of the line of sight by a foreground dust
  halo is included in the \hetg, but not \suz, fit.  Residuals
  (histogram colors match their respective data colors) show the fit
  absent both the Fe\,K$\alpha$ emission and ionized absorption
  \protect{\citep{nowak:11a}}.  Right: the same spectra as on the
  left, now with the addition of quasi-simultaneous \epic spectra
  (orange stars), fit with an absorbed and dust-scattered
  (\chandra-\hetg and \epic only) powerlaw (model not shown), with
  cross normalizations set so that the spectra match at 7\,keV.}
\label{fig:spectra}
\end{figure*}

As a first example, Fig.~\ref{fig:spectra} shows a spectrum of the
black hole candidate (BHC) and X-ray binary, \cyg.  The observation
occurred at orbital phase 0 (superior conjunction), when the optical
companion was directly between our line of sight to the black hole.
This particular observation was quasi-simultaneous in \emph{all X-ray
  satellites flying at that time} (April 2006), with the
\chandra-\hetg and \suz spectra previously being discussed in
\citet{nowak:11a}.  There are a number of issues that arose in
analyzing these spectra.  First is the effect of spatial resolution in
\xmm and \chandra.  To a first approximation, the presence of
foreground dust scatters soft X-rays out of the arcsec scale field of
view of these instruments, but X-rays scatter back in (albeit
time-delayed) on the arcminute scale field of view of instruments such
as \suz.  This effect is noticed on the right side of
Fig.~\ref{fig:spectra}, where the ``flux-corrected'' spectra from
\epic and \hetg lie below that of \suz.  This is \emph{not} (or at
least not predominantly) a calibration effect, but is instead due to
dust scattering that had to be included in the modeling on the left
portion of Fig.~\ref{fig:spectra}. This effect is described in more
detail for various X-ray instruments by \citet{corrales:16a}.

A recent \chandra/\swift study of the black hole candidate V404~Cyg by
\citet{heinz:16a} shows that observations of black holes with
long-term time variations can probe the location and composition of
dust in the interstellar medium (ISM). \xmm is perhaps the current
best instrument to perform such observations going forward, as it has
the proper combination of high (enough) spatial resolution with large
effective area in the soft X-rays, yet is relatively free of pileup
(see \citealt{davis:01a}) compared to \chandra.

\begin{figure*}
\begin{center}
\includegraphics[width=0.59\textwidth,viewport=0 0 593 287]{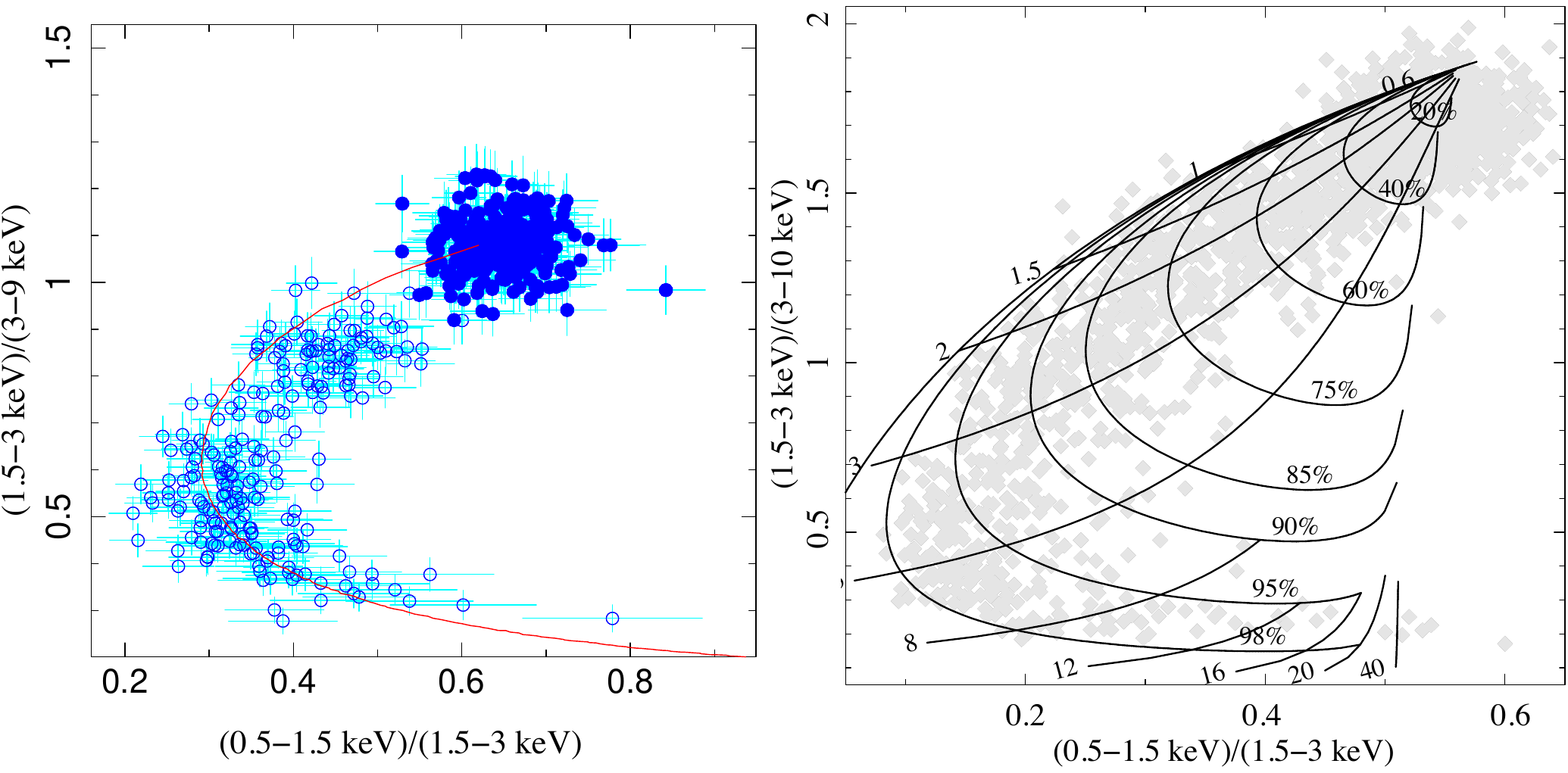}
\includegraphics[width=0.37\textwidth,viewport=84 20 585 384]{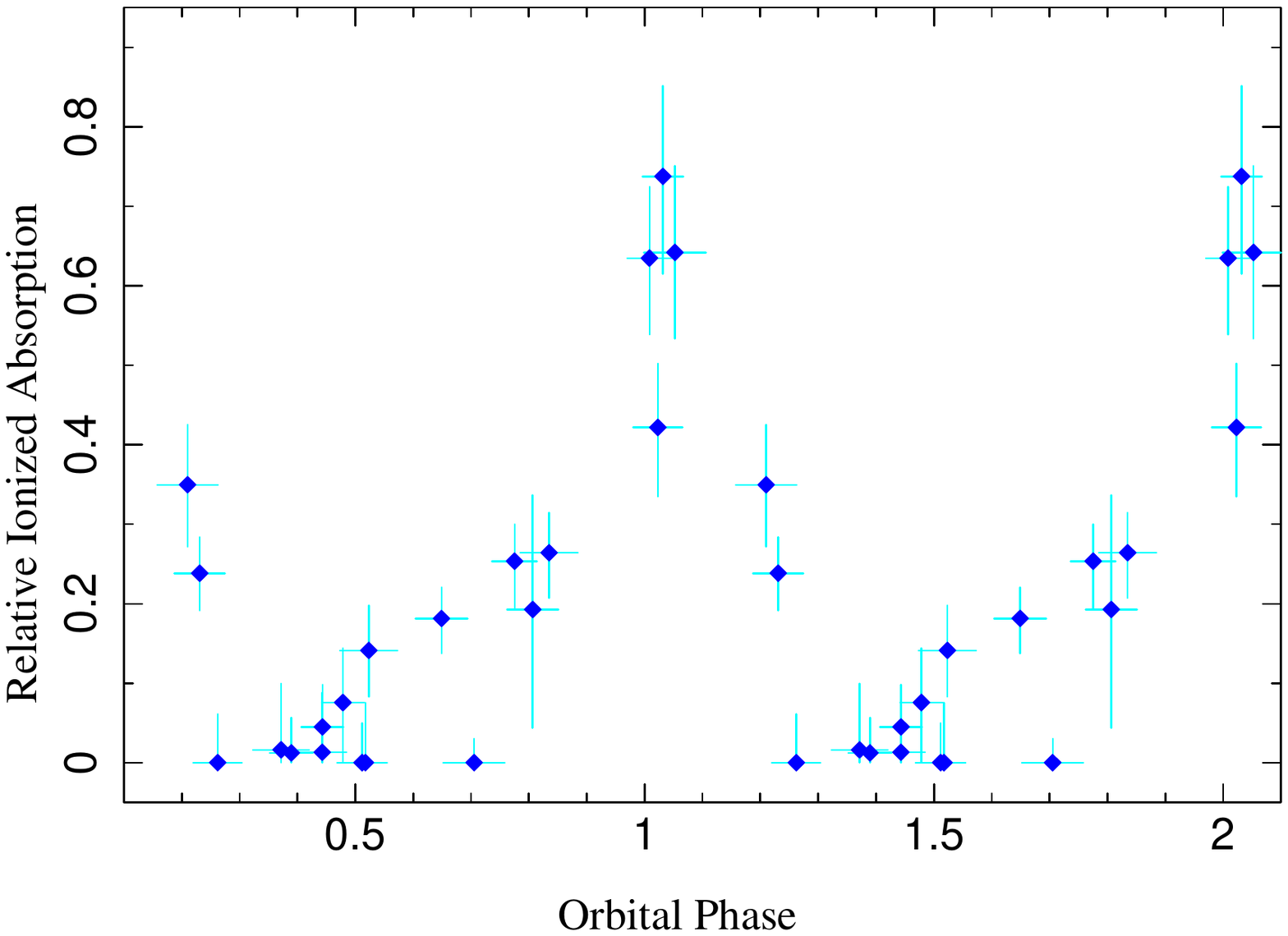}
\end{center}
\vspace{-0.05 in}
\caption{Left: Color-color diagram of the full \suz observation of
  \cyg, from Fig.~\ref{fig:spectra}, now including dipping events.
  The red line is a simple partial covering model that describes these
  events with a varying column, but fixed partial covering fraction
  \protect{\citep{nowak:11a}}.  Middle: Color-color diagram derived
  from \chandra-\hetg data of \cyg, including dipping events.  Lines
  show simple partial covering models as on the left, with covering
  fractions ranging from 20--98\% and columns ranging over
  (0.6--40)$\times10^{22}\,\textrm{cm}^{-2}$.  (See the discussions in
  \citealt{hanke:09a}.)  Right: Relative degree of ionized absorption
  in \suz observations of \cyg hard states, compared to that found in
  the orbital phase 0 (superior conjunction) observation of
  Fig.~\protect{\ref{fig:spectra}}.  All of these fits are for
  \emph{times without any dipping events}.}
\label{fig:color}
\end{figure*}

The second effect noticeable in Fig.~\ref{fig:spectra} is the presence
of ionized absorption, clearly visible in the \chandra-\hetg spectra,
yet completely unresolved but \emph{extremely statistically
  significant} in the \suz and \epic spectra \citep{nowak:11a}.  It is
important to note that both the ionized absorption, as well as
prominent dipping events that occur as a function of orbital phase in
\cyg, are associated with absorption by a powerful wind from the O
star secondary in this system (see the discussions and references in
\citealt{hanke:09a,misk:16a}).  The spectra shown here \emph{occur
  outside of all detectable dipping events at phase 0}.  Such ionized
absorption, only spectroscopically resolvable by \rgs, \letg, and \hetg
spectra, is a ubiquitous feature in spectra of \cyg \citep{misk:16a}.
This is demonstrated in the right most figure of Fig.~\ref{fig:color},
where the relative amplitude of this component is shown as a function
of orbital phase in \emph{non-dip} \suz spectra of \cyg.

Fig.~\ref{fig:color} shows color-color diagrams from the full \suz and
\chandra observations associated with Fig.~\ref{fig:spectra}.  The
extension to the lower left (excluded in the shown spectra) is due to
the interspersal of dense clumps passing in front of our line of sight
during the observations.  The ``tail'' from the lower left to the
lower right, however, is due to the ``partial covering'' nature of
these clumps.  In \suz, this is also an artifact of the dust
scattering halo, as we are seeing the time-delayed scattering from
foreground halos on arcminute scales, with the fitted covering
fraction being consistent with the fraction of flux scattered by dust
\citep{nowak:11a}.  Such scales are resolved out by the \chandra point
spread function (PSF); however, there remains an $\approx 2\%$
uncovered fraction.  Although this may be due to the inner core of the
dust halo, another possibility is that it is due to partial covering
in time.  That is, we might be seeing large dipping events passing
through our line of sight on time scales faster than the $\approx
1.8$\,s integration times of these \chandra observations.  Exploring
this possibility, however, would require the larger effective area and
faster timing capabilities of \textsl{XMM}.

\begin{figure*}
\begin{center}
\includegraphics[width=0.46\textwidth,viewport=25 0 578 256]{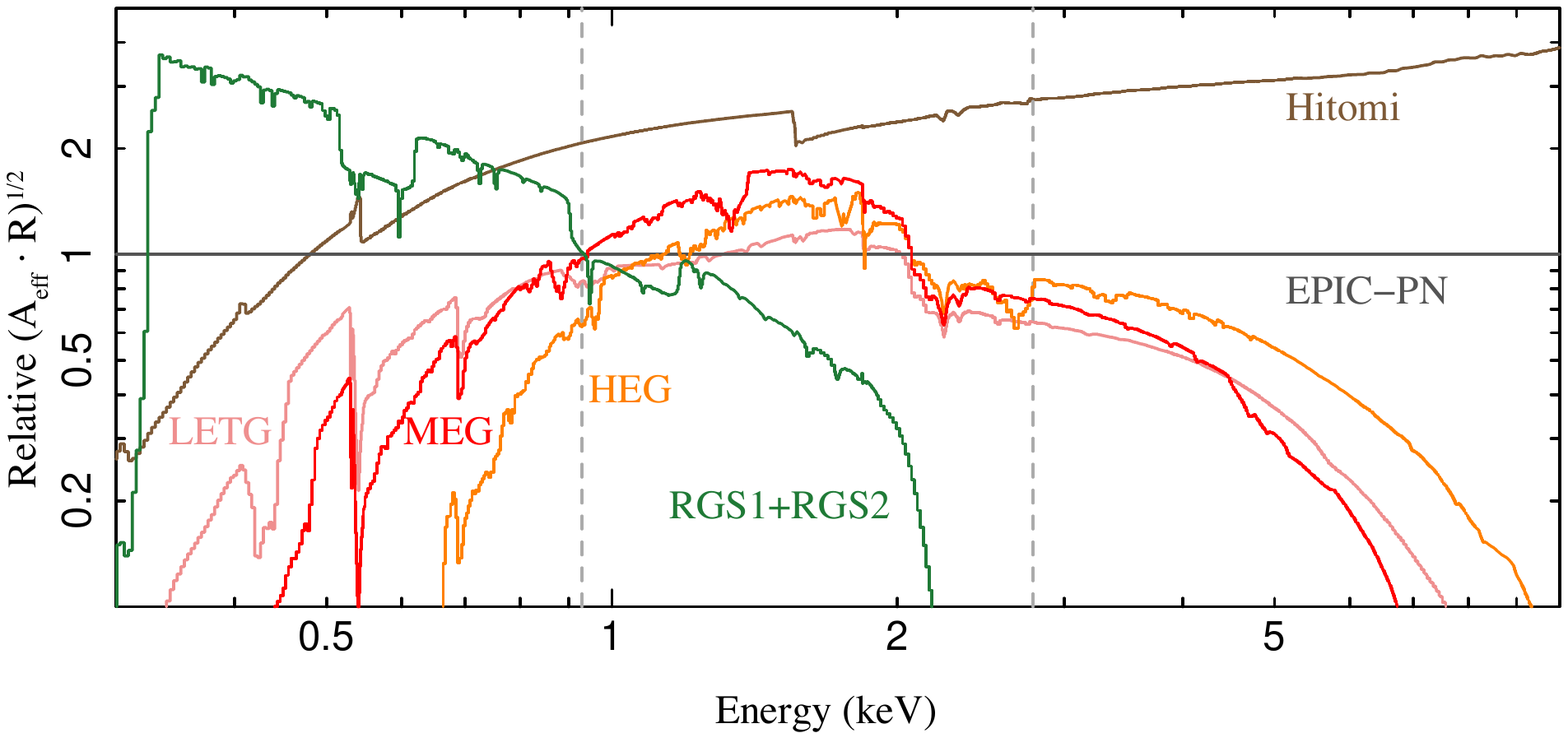}
\includegraphics[width=0.26\textwidth,viewport=5 0 395 325]{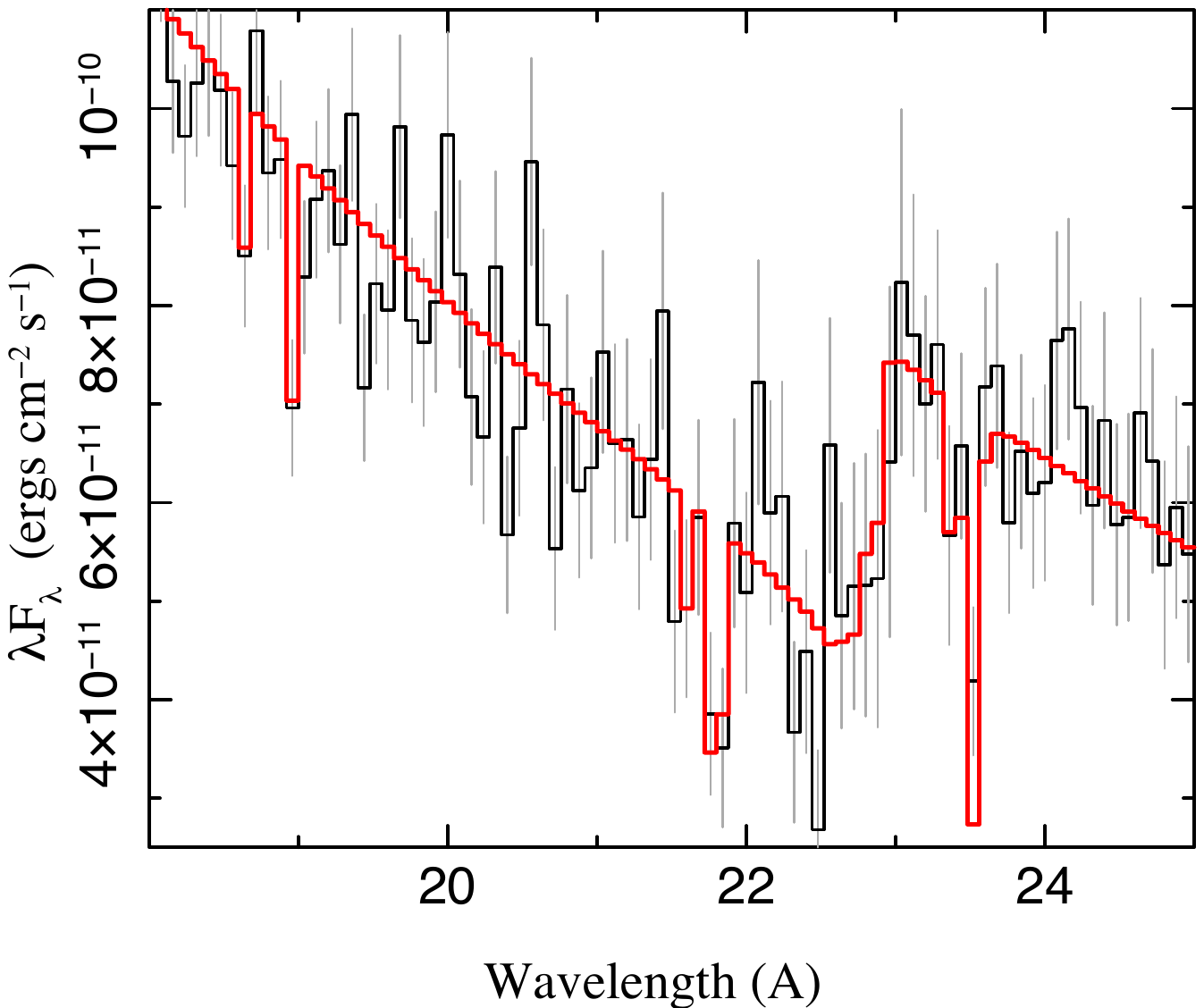}
\includegraphics[width=0.26\textwidth,viewport=5 0 395 325]{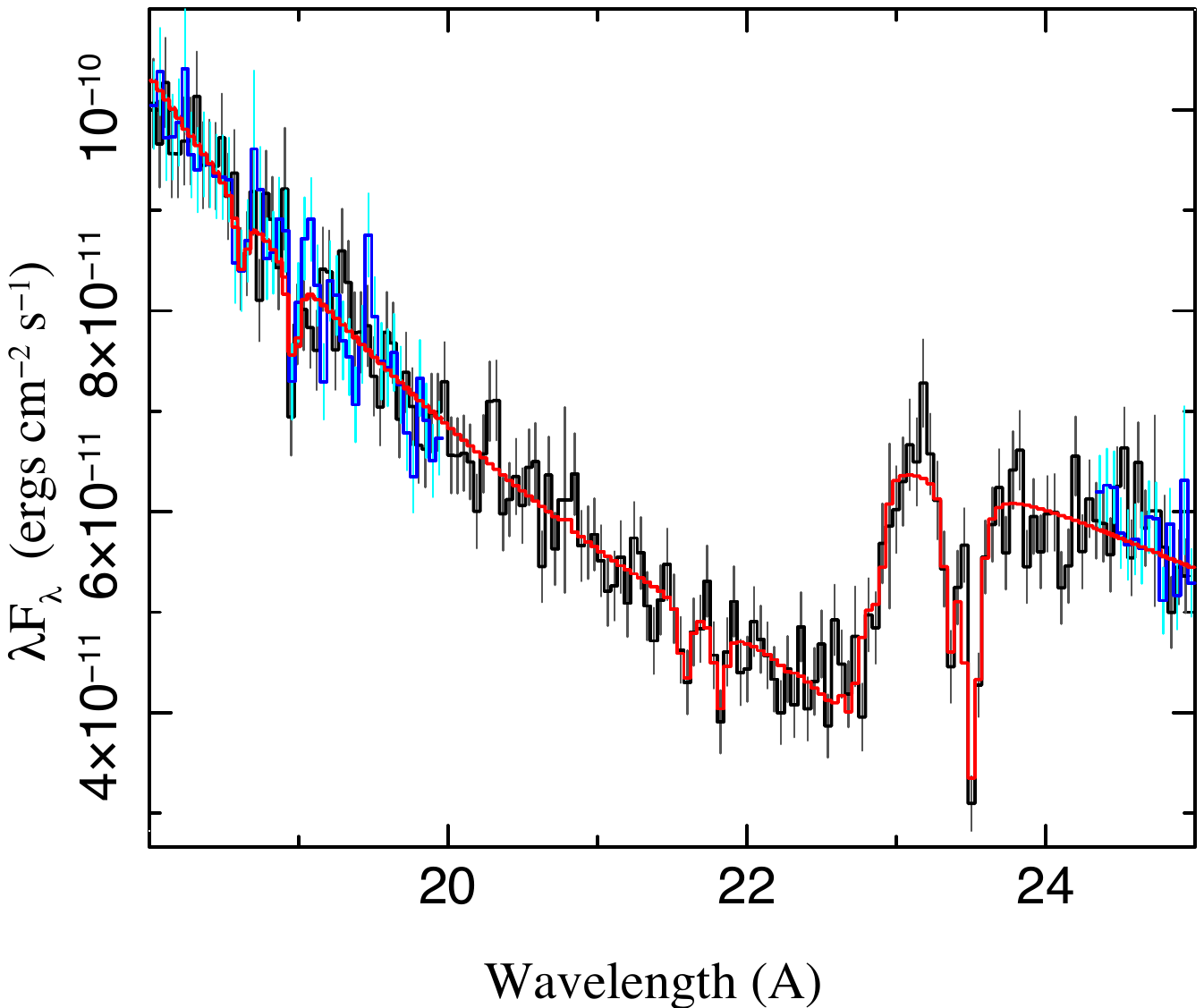}
\end{center}
\vspace{-0.1 in}
\caption{Left: Figure of merit for emission and absorption line studies 
  (square root of effective area times spectral resolution) for
  various instruments, normalized to that for \epic.  The vertical
  dashed lines show the energies below which \chandra-\heg (right) and
  \xmm-\rgs (left) spectral resolution exceeds that achieved by the
  \hitomi calorimeter.  Middle/Right: Comparison of two comparable (in
  terms of flux, spectral shape, and exposure time) \hetg (middle) and
  \rgs (right) observations of the oxygen-edge region of the black
  hole candidate \fu (see \protect{\citealt{nowak:11b}}).  The
  spectrum is a multi-color disk with peak temperature $\approx
  1.7$\,keV absorbed by a neutral column of
  $1.2\times10^{21}\,\textrm{cm}^{-2}$.  There is also evidence for
  ionized absorption by the warms phase of the interstellar medium, as
  well as by material local to the system.}
\label{fig:arfs}
\end{figure*}

This latter experiment potentially has been conducted with the recent
(Summer 2016) observations of \cyg conducted by \xmm, which were
designed (PI: P. Uttley) to conduct spectral-timing studies of the
continuum, Fe band, and reflection spectra of \cyg, as well as to
perform high-resolution absorption spectroscopy of the secondary
wind. It is important to note here that \xmm-\rgs is ideally suited
for such studies as it has the highest ``figure of merit'' for soft
X-ray band spectroscopy, as shown in Fig.~\ref{fig:arfs}.  \emph{This
  would still be true below $\approx 0.8$\,keV, even if the \hitomi
  mission had not been lost.}  The power of this \rgs capability can
be seen in Fig.~\ref{fig:arfs} where two (non-simultaneous, but very
comparable in terms of flux, spectral shape, and integration time)
observations of the same BHC are shown, highlighting \meg and \rgs
spectra of the oxygen absorption edge region.  Among potential future
studies to be considered with \rgs are detailed spectral-temporal
observations of galactic black hole systems, especially over binary
orbits, to separate out absorption components local to the system from
those due to the ISM.

The advantage of galactic binary BHC studies, aside from spectral
signal-to-noise, are the time scales involved.  In \cyg, thanks to
high resolution spectroscopy from \rgs and \hetg, we know that both
ionized absorption and ``partial covering'' models are applicable.  We
can track their changes over binary orbits, and thus have the strong
promise of separating out their effects from those associated with the
inner accretion flow, e.g., the relativistically broadened
Fe~K$\alpha$ line.  Such effects are more difficult, but no less
important, to disentangle with high resolution spectroscopic
observations of Active Galactic Nuclei (AGN).  This is directly
relevant to studies of relativistic features in AGN, where some
researchers have suggested that what is modeled as a broadened line is
in fact due to features dominated by partial covering
\citep{miller:09a}.  For the case of NGC~3783, \citet{brenneman:11a}
showed the necessity of simultaneously fitting the warm absorption and
broad line features.  Further work by \citet{reynolds:12a} showed how
the model parameter regions of \citet{miller:09a} and
\citet{brenneman:11a} are related, with the latter being favored in
Markov Chain Monte Carlo (MCMC) analyses of the spectra.  Also
important in the work of \citet{brenneman:11a} and
\citet{reynolds:12a} was the inclusion of hard X-ray spectra, which we
discuss further below.

\section{\xmm/\chandra Synergies}\label{sec:synergy}

The figure of merit presented in Fig~\ref{fig:arfs} tells only part of
the story; absolute spectral resolution is still an important metric.
For the absorption edge of \fu shown in Fig.~\ref{fig:arfs}, the
higher effective area of the \rgs as compared to the \meg clearly
greatly benefits our ability to model these features.  Here, however,
the edge and its associated absorption lines (i.e., transitions of
atomic oxygen; see \citealt{juett:04a}), as well as the possible
ionized features from either the interstellar medium (see
\citealt{yao:05a}) or the local system \citep{nowak:08a}, are
well-isolated.  This is not always the case, especially with
absorption features in the soft X-ray band covered by the \rgs.

\begin{figure}
\begin{center}
\includegraphics[width=0.42\textwidth,viewport=30 15 565 755]{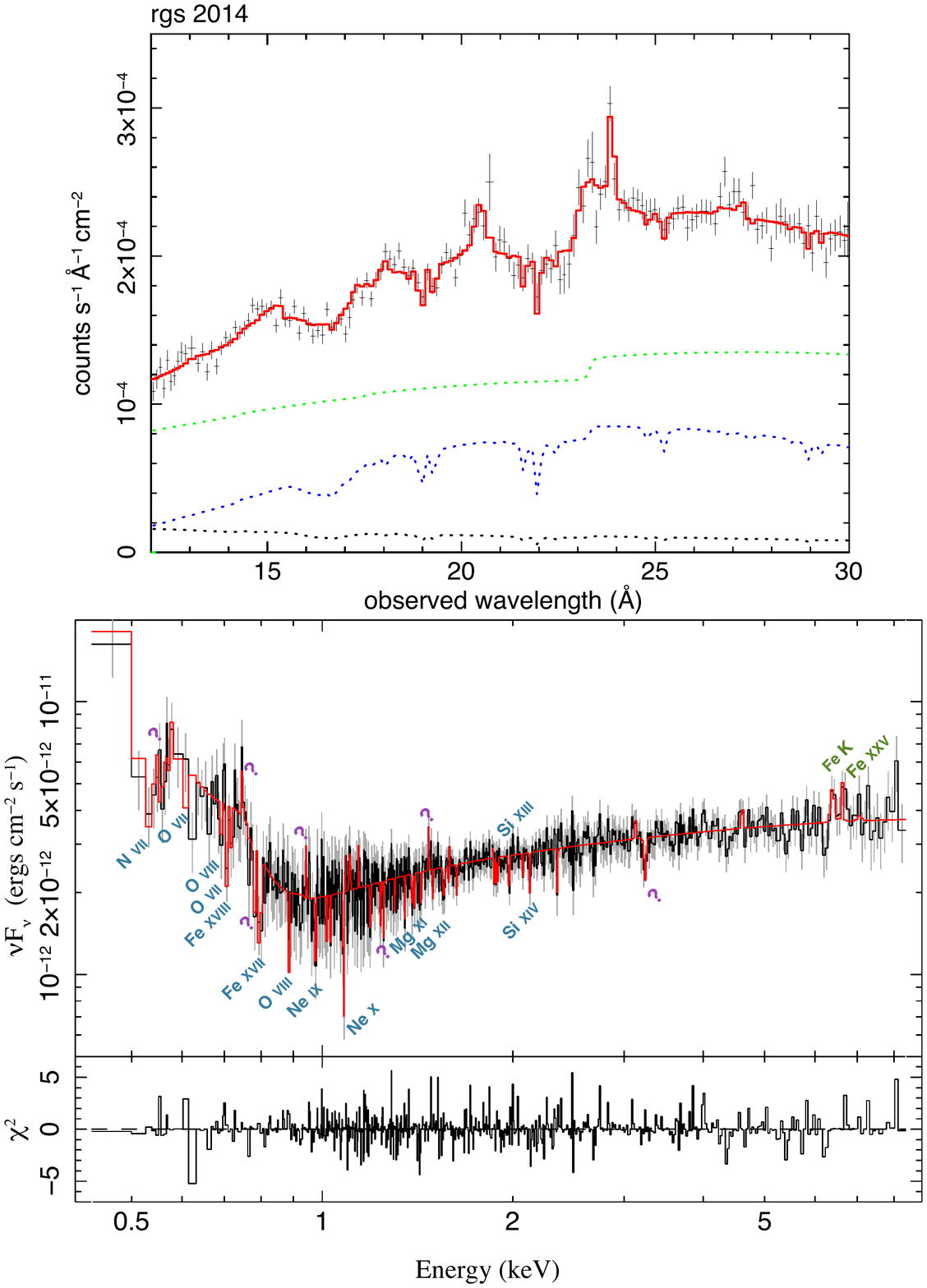}
\end{center}
\vspace{-0.1 in}
\caption{Top: \xmm-\rgs spectra of the Seyfert galaxy \pg (figure from
  \protect{\citealt{pounds:16a}}; see also
  \protect{\citealt{pounds:16b}}), fit with an absorbed continuum and
  several ionized outflow components.  Bottom: \chandra-\hetg spectrum
  of \pg (Danehkar et al. 2016, in prep.).  The figure is labeled with
  possible lines, found from a blind search, consistent with an
  outflowing wind at a single (galaxy restframe) blueshift.  Question
  marks are potential lines also found in the blind search that have
  not yet been identified with absorption or emission components.}
\label{fig:pg}
\end{figure}

As an example in Fig.~\ref{fig:pg} we show a portion of an \rgs
spectrum of the Seyfert galaxy \pg, from the work of
\citet{pounds:16a,pounds:16b} (see also Pounds et al. 2017, in this
volume).  Based upon both prior \epic studies \citep{pounds:03a} and
these more recent \rgs studies, it has been argued that \pg exhibits
blueshifted, ionized outflows moving at speeds in the galaxy rest
frame of ${\cal O}(10\%)$ the speed of light.  The best evidence,
however, is that there are \emph{multiple} ionized components at a
variety of outflow velocities that are blended in the \rgs spectra
\citep{pounds:16a,pounds:16b}.  Furthermore, basic line properties,
such as line widths, are not resolved by the \rgs spectra. Cleanly
separating ionized outflow components and resolving their individual
line widths requires the resolution of the \chandra-\hetg; however,
such observations require large investments of observing time.

To this end, we have obtained an \hetg observation (PI: Julia Lee) of
\pg for a total integration time of 450\,ks.  Fig.~\ref{fig:pg} shows
the binned, combined \hetg spectra, with ``blind search'' line fits
(Danehkar et al. 2016, in prep.) derived from the unbinned
spectra. This figure further shows possible line identifications for a
set of features at a common blueshift of $0.06c$, consistent with one
of the outflows described by \citet{pounds:16b}.  The full analysis
will be described elsewhere by Danehkar et al. (2016); however, we
note that the analysis heavily draws upon the higher effective area,
albeit lower resolution, \rgs analysis.  It likely would have been
impossible to obtain the \hetg observations for this integration time,
or have found these features in even a blind search in shorter \hetg
observations, without first having the preliminary studies with
\xmm-\rgs.

\begin{figure}
\begin{center}
\includegraphics[width=0.48\textwidth,viewport=85 25 580 382]{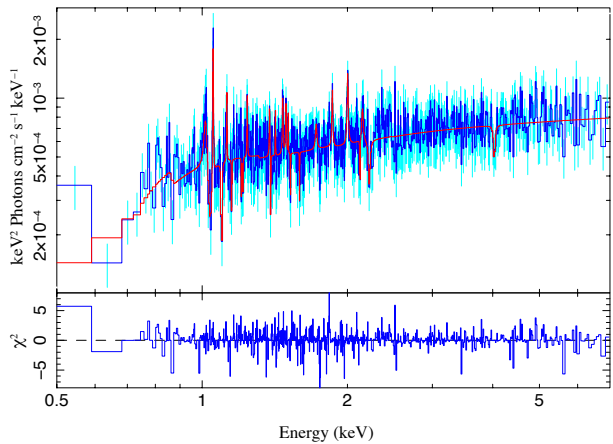}
\end{center}
\vspace{-0.1 in}
\caption{Simulated 500\,ksec \chandra-\hetg spectrum of the ULX NGC
  1313 X-1, based upon ionized absorption and emission line fits of
  \rgs spectra presented by \citet{pinto:16a} (see their Fig.~1).  The
  model fit corresponds to a ``blind line search'' implemented in
  \isis \protect{\citep{houck:00a}} (see \S\protect{\ref{sec:soft}}),
  and recovers many of the features from the input model.}
\label{fig:ulx}
\end{figure}

A further example of this synergy between \xmm-\rgs and \chandra-\hetg
is shown in Fig.~\ref{fig:ulx}, where we show a simulation of an
upcoming 500\,ks \hetg observation of NGC~1313 X-1. Analysis of
archival \rgs observations suggests ionized emission at systemic
velocities, as well as ionized absorption at both systemic and
blueshifted outflow ($v\approx0.1c$) velocities \citep{pinto:16a}.
These components, however, are unresolved by the \rgs, but would have
taken significant integration times to discover with \hetg
observations alone.  Based upon the initial \rgs studies, a commitment
was made to use 500\,ks (out of an annual allocation of 700\,ks to
\hetg PI Claude Canizares) of \hetg Guaranteed Time Observations
(GTO).  The goals of these observations will be to resolve line widths
and to obtain more precise outflow velocity measurements.  Comparing
the simulations shown in Fig.~\ref{fig:ulx} to Fig.~\ref{fig:arfs}, we
see that much of the \hetg gains will come from the 1--2\,keV region
where its figure of merit exceeds that of \rgs.  We see again,
however, the utility of having first used the higher effective area
\rgs to identify a high resolution spectroscopy target of interest.
Furthermore, the \rgs observations allowed us to determine a plausible,
albeit significant, \hetg integration time for follow-up study.

\section{Line Analysis Software}\label{sec:soft}

A difficulty in the analysis of high resolution X-ray spectra is the
complexity of models required to describe the data.  Above $\approx
3$\,keV, e.g., the Fe region, where spectral lines are relatively few
and widely spaced, simple direct modeling of individual line features
may be feasible.  In the soft X-ray bands covered by the \rgs, \letg,
and \meg, lines can be numerous and closely spaced, complicating the
analysis.  Few codes exist for describing emission and absorption by
the photoionized plasmas relevant to the study of black hole systems,
among them are \cloudy, \citep{ferland:98a,ferland:13a}, \xstar
\citep{kallman:01a}, and the models included in the \spex analysis
package \citep{kaastra:96a}.  These codes take a global approach to
description of the high resolution spectra, modeling multiple
individual lines and blends simultaneously.  As such, however,
searching parameter space can be slow and unintuitive, and the codes
and code results can be difficult to interpret.  E.g., for the lines
that comprise the best fit, identification of individual components,
and obtaining access to the relevant atomic data that led to those
components, can be difficult.  Creating custom interfaces to these
codes can be cumbersome, especially in cases where the source code is
not publicly available (e.g., \spex).  These issues can lead to a
``potential barrier'', discouraging the use of high resolution X-ray
spectra.

To further reduce the barrier to entry to the study of high resolution
X-ray spectra, a number of researchers have been creating more
user-friendly tools for both fitting and interpreting such models.
Graphical user interfaces to the atomic databases for piecewise fits
to restricted wavelength regions are being explored.  (See the
individual talks by T. Kallman and R. Smith at the 2016
meeting\footnote{\texttt{web.mit.edu/iachec/meetings/2016/}} of the
International Astronomical Consortium for High Energy
Calibration[IACHEC]).  \xstardb is a code suite for use within \isis
that provides easier access to the \xstar atomic data (e.g., individual line
identifications, line searches based on transition strengths, etc.), that
also manages and runs \xstar spectral fits.

In the example of \pg above, we have employed an \isis script suite
that we are developing that allows for more phenomenological analysis
of high resolution data.  (Preliminary code versions are available as
part of the \isis scripts maintained and developed at Remeis
Observatory\footnote{\texttt{www.sternwarte.uni-erlangen.de/isis/}}.)
The scripts allow for modeling of individual lines using a choice of
standard profiles (e.g., gaussian or Voigt profiles, with parameters
input in energy or wavelength, with or with explicit redshift
parameters).  Lines are added to the model parameter file either in
wavelength or energy order, with users being able to name the line
profile (e.g., a redshifted \texttt{gaussian} added as a description
of Fe K$\alpha$ could be named \texttt{zg\_FeKa}).  For
Fig.~\ref{fig:pg}, the lines were placed in a multiplicative model (to
allow them to describe either emission or absorption lines, with the
latter never formally describing negative counts), with their names
enumerating the statistical order in which they were added in a blind
line search.  (E.g., the most significant line, likely from Ne{\sc
  x}, was called \texttt{zg\_0}, while the seventh most significant
line, likely from Ne{\sc ix}, was called \texttt{zg\_6}.)

We plan to expand this code to allow for the easy addition of multiple
levels of functional ties, such as tying redshifts across profiles
expected to come from similar temperature/ionization states, while
simultaneously tying line strengths within a line series.  The scripts
will include procedures for applying blind line searches to the data.
The goal is to allow a preliminary, straightforward phenomenological
description of the spectra as a prior step before embarking upon more
physical descriptions with, for example, complex and slow running
photoionization codes.

\section{Summary}\label{sec:sum}

As discussed above, high resolution spectra provide unique information
about black hole systems.  Looking forward to future observations with
\xmm-\rgs and \chandra-\hetg, one important area that we must consider
is broadband, multi-satellite observations.  For the example of
NGC~3783, as shown by \citet{brenneman:11a} and \citet{reynolds:12a},
both high resolution, soft X-ray spectra and broadband (in that case,
\suz; going forward in the future, most likely utilizing \nustar) are
necessary to understand relativistic features.  Indeed,
\citet{garcia:15a} has recently shown that under ideal circumstance,
given knowledge of the underlying continuum and quality observations,
the soft X-ray reflection spectrum \emph{measures} the cutoff at high
energies.  In practice, however, concerns about ionized absorption, as
discussed in \citet{nowak:11a} for the case of \cyg, must first be
addressed with, e.g, high spectral resolution observations.  Clearly,
however, just as their is a synergy between CCD and gratings quality
spectra (\S\ref{sec:lowhi}), and a synergy between large effective
area \rgs spectra and higher resolution \hetg spectra
(\S\ref{sec:synergy}), there is also a synergy between soft X-ray and
hard X-ray spectroscopy.

It is somewhat unfortunate, then, that this avenue has not yet been
strongly explored.  Although there have been many successful
simultaneous \xmm and \nustar observations, an examination of the
literature to date shows that of the approximately 100 papers
describing \xmm/\nustar observations, only 11 discuss \rgs spectra in
any capacity.  Of the dozen accepted joint \chandra/\nustar proposals,
only three utilize the \hetg.  (One of these campaigns is a joint
observation of NGC~3783; PI L. Brenneman.)  

When sufficient telemetry exists, and instrumental constraints aren't
violated, \rgs spectra will always be obtained with an \xmm
observation.  This does not necessarily guarantee that the \rgs will
have sufficient signal-to-noise for use, but in many cases it does.
How do we encourage users to analyze these data?  We note that the work
of \citet{pinto:16a} was derived from archival observations from
proposals that were initially designed to obtain CCD-quality spectra.
How do we encourage users to explicitly design programs around the use
of \rgs spectra?

For \chandra proposals, gratings observations (typically dominated by
the use of \hetg) comprise only $\approx 15\%$ of the accepted
program.  This is not because gratings proposals are accepted at a
lower rate than non-gratings proposals, rather it is because they 
represent only 15\% of the submitted proposals.  How do we encourage
users to apply for more gratings observations?

Part of the issue is undoubtedly the complexity of the spectra, and
the lack of an ``easy entry'' to the high resolution spectroscopy
software (\S\ref{sec:soft}).  The high resolution spectroscopy
community must continue to develop tools that allow for wider use of
existing and future data.  Examples--- such as those in
Fig.~\ref{fig:spectra} or the case of NGC~3783 and similar AGN---
where CCD-quality spectra cannot be properly analyzed without
knowledge of the high resolution spectra must also be emphasized.

As shown in Fig.~\ref{fig:arfs}, \rgs below $\approx0.9$\,keV and
\hetg between $\approx 0.9$--2\,keV are the premiere instruments for
spectroscopic studies.  A relaunch of a satellite comparable to
\hitomi would not alter the situation for \rgs.  High resolution
spectroscopy has the capability of providing measures of density,
temperatures, and velocities, and is in fact the prime scientific
mission of the proposed \athena and \xrs missions.  Continued studies
with \rgs, \letg, and \hetg must pave the way for these future
missions.

\acknowledgements

Michael Nowak gratefully acknowledges funding support from the
National Aeronautics and Space Administration through the Smithsonian
Astrophysical Observatory contract SV3-73016 to MIT for support of the
\chandra X-ray Center, which is operated by the Smithsonian
Astrophysical Observatory for and on behalf of the National
Aeronautics Space Administration under contract NAS8-03060.  He
further acknowledges support by NASA Grant NNX12AE37G. He would like
to thank V. Grinberg, D. Huenemoerder, M. Middleton, and J. Wilms for
useful conversations.

\newpage



\end{document}